# AutoLoop: a novel autoregressive deep learning method for protein loop prediction with high accuracy

Tianyue Wang[1,2#], Xujun Zhang[1#], Langcheng Wang[3], Odin Zhang[1], Jike Wang[1], Ercheng Wang[4], Jialu Wu[1], Renling Hu[1], Jingxuan Ge[1], Shimeng Li[1], Qun Su[1], Jiajun Yu[2,5], Chang-Yu Hsieh[1,*], Tingjun Hou[1,*], Yu Kang[1,2*]

[1]College of Pharmaceutical Sciences, Zhejiang University, Hangzhou 310058, Zhejiang, China

[2]Shanghai Innovation Institute, Shanghai 200231, China

[3]Department of Pathology, New York University Medical Center, 550 First Avenue, New York, NY 10016, USA

[4]Zhejiang Laboratory, Hangzhou 311100, Zhejiang, China

[5] College of Computer Science and Technology, Zhejiang University, Hangzhou, China

[#]Equivalent authors

***Corresponding authors**

**Yu Kang**

**Email:** yukang@zju.edu.cn

**Tingjun Hou**

**E-mail:** tingjunhou@zju.edu.cn

**Chang-Yu Hsieh**

**Email:** kimhsieh2@gmail.com

## Abstract

Protein structure prediction is a critical and longstanding challenge in biology, garnering widespread interest due to its significance in understanding biological processes. A particular area of focus is the prediction of missing loops in proteins, which are vital in determining protein function and activity. To address this challenge, we propose AutoLoop, a novel computational model designed to automatically generate accurate loop backbone conformations that closely resemble their natural structures. Uniquely, AutoLoop employs a bidirectional training approach while merging atom and residue level embedding, thus bolstering its robustness and precision. To validate its efficacy, we compared AutoLoop with twelve established methods, including FREAD, NGK, AlphaFold2 and AlphaFold3. The results indicate that AutoLoop consistently outperforms other methods, achieving a median RMSD of 1.12 Å and a 2 Å-success rate of 73.23% on the CASP15 dataset, also maintaining its superior performance on the HOMSTARD dataset. Notably, AutoLoop demonstrates the best performance across almost all loop lengths and secondary structural types. Beyond its accuracy, AutoLoop's computational efficiency is also remarkable with an average processing time of 0.10s per generation. The addition of a post-processing module (i.e. side-chain packing and energy minimization) enhances AutoLoop's performance slightly, reflecting the sound reliability of the predicted backbone structures. Additionally, the case study exhibits a certain potential of AutoLoop for precise predictions based on one or several dominant loop conformations. These advancements hold great promise for protein engineering and drug discovery, paving new ways for designing more potent therapeutic agents.

## Introduction

Protein is the basement of life processes, participating in numerous biological functions. A comprehensive understanding of protein structures underpins several downstream applications such as drug design and mechanism exploration[1]. However, more than half of protein contains the missing region in Protein Data Bank[2] (PDB) which usually consists of loop[3]. These loop regions are usually located on the surface of the protein, and serve as connectors between $\alpha$-helix and $\beta$-sheets[4, 5]. They are typically highly flexible and play important roles in processes such as molecular recognition, enzyme catalysis, allosteric regulation or signaling[6-9]. It has been widely acknowledged that the high flexibility of loops rendered them challenging in protein structure prediction[10]. Therefore, addressing the missing loop in protein structures requires advanced computational methods.

Previous computational tools involved in predicting loop structures can be divided into three categories: knowledge-based, *ab initio* and hybrid methods. Knowledge-based methods, such as FREAD[11], LoopIng[12], Prime[13, 14], ArchPred[15] and DaReUS-Loop[16], typically utilize a template database alongside various search algorithms. While these methods are computationally efficient, their performance is often limited by the quality and breadth of the template databases they rely on. *Ab initio* methods, such as Next generation KIC (NGK)[17], GalaxyLoop-PS2[18] and Distance-guided Sequential Chain-Growth Monte Carlo ($D_ISG_{RO}$[4]) generally encompass a two-stage process involving conformation sampling and scoring. These methods facilitate the prediction of protein structures independent of template databases, thereby allowing for an extensive exploration of conformational space[19]. However, these methods typically require more computational resources compared to the knowledge-based approaches, so they become impractical as the size of the conformational space increases exponentially with the length of the protein loop[20]. Hybrid loop modeling techniques combine elements of both *ab initio* and knowledge-based methods to improve the accuracy of protein loop structure predictions.

The advancement of Artificial Intelligence (AI) has offered new possibilities to protein engineering[21, 22] and the following drug discovery[23-26]. Recently, the scientific community witnessed a significant advancement with the introduction of AlphaFold2[27], which demonstrated superior performance at the 14th Critical Assessment of protein Structure Prediction (CASP14) event, and was considered to largely solve the 50-year-old protein folding problem[19]. RoseTTAFold[28] was proposed concurrently, which made broad predictions on the protein structure accessible. In a more recent development, D-I-TASSER emerged as the top-performing server in both the Single-domain and Multi-domain categories at CASP15, according to its website (https://zhanggroup.org/D-I-TASSER/). Building on previous achievements, AlphaFold3[29] has further refined the predictive models for complex interactions, such as those between proteins. Despite these advancements, several studies including our previous work have highlighted that both AlphaFold2 and RoseTTAFold exhibit limitations in accurately predicting protein loop regions[20, 30, 31]. Nguyen et al.[32] have then developed a deep-learning approach for reconstructing protein loops. Unfortunately, the absence of openly available code for this method poses constraints on its broader application and verification by the scientific community.

In our previous work[20], we assessed thirteen well-known methods, including traditional approaches (knowledge-based and *ab initio* methods) and deep learning techniques (AlphaFold2[27] and RoseTTAFold[28]), across two benchmark datasets of over 10,000 samples. The results indicated that FREAD exhibited the best performance, which is a knowledge-based method that searches candidates by utilizing the distance of anchor C$\alpha$ atoms and filters them using environment substitution scores, statistical energy functions, and anchor root mean square deviation (RMSD). NGK and RML[33] outperformed other *ab initio* methods, employing techniques such as taboo sampling and Ramachandran distribution-based phi/psi sampling, annealing with Ramp repulsive and Ramp rama. AlphaFold2 surpassed RoseTTAFold in loop prediction across many scenarios[20, 30]. Additionally, $D_ISG_{RO}$ and Prime demonstrated high computational efficiency. Generally, the accuracy and efficiency of these methods for loop structure prediction still need improvement[20]. Besides, some methods[34, 35] cannot be implemented locally, which hinders its further applications.

In this study, we introduce an advanced end-to-end deep learning approach AutoLoop (Figure 1) specifically designed for protein loop prediction. AutoLoop distinguishes itself with high accuracy and efficiency by predicting loop backbone conformation that closely resembles the native structure of proteins. This typically requires extensive exploration and filtering of the conformational space, a process known for being time-consuming. AutoLoop, however, leverages an autoregressive model to significantly reduce the size of this conformational space and simplify the selection process for users, allowing them to choose the intended conformation more effectively and reduce the potential for errors that may arise from imprecise scoring functions. Furthermore, AutoLoop offers a novel solution to a common issue encountered in previous methods, where changes to the loop fragment could hinder proper closure at the C-terminus.[36-39] It employs a bidirectional training strategy that systematically predicts the loop conformation prediction module from the N-terminal to the C-terminal and vice versa. This technique effectively addresses the limitations of earlier models, enhancing the understanding and accurate prediction of the dominant loop backbone conformation. Post-processing steps including side-chain packing and energy minimization can be easily integrated into AutoLoop, thereby solidifying its role as a useful tool for accurate loop prediction and further applications.

Based on our previous assessments[20], the notably effective methods, namely NGK, Prime, D$_{I}$SG$_{RO,}$ FREAD, RML, AlphaFold2, AlphaFold3, RoseTTAFold, and ColabFold[40] were chosen for further comparison against AutoLoop using the most recent CASP15 competition data. Moreover, for a thorough validation, NGK and a series of web-server-provided methods (i.e., LoopIng, DaReUS-Loop, Galaxy PS2, and Sphinx) were tested on another commonly used benchmark dataset HOMSTARD[41]. This dataset contains only a limited number of samples that are testable exclusively through these web-server-provided methods. According to the CASP15 results, AutoLoop exhibits the best performance, with median backbone RMSD as 1.12 Angstrom (Å), average RMSD as 1.90 Å and 2 Å, 1 Å success rate of 73.23% and 48.92%, respectively. Following the post-processing (side-chain packing and energy minimization) the median, average RMSD decreases to 1.10 Å, 1.53 Å, and the success rates increase to 79.19% and 44.41% for the 2 Å and 1 Å thresholds, respectively. Moreover, on the HOMSTARD test dataset,

AutoLoop achieves median and average RMSD of 2.34 Å and 2.16 Å, respectively. After post-processing, these values reduced to 1.47 Å and 1.39 Å, respectively, all of which also significantly surpass those of the second-best performing methods. AutoLoop also demonstrates superior performance across various loop lengths and secondary structural types in the CASP15 dataset, highlighting its robust capability to accurately predict loops of differing lengths and secondary structural types. Regarding computational efficiency, AutoLoop completes tasks with notable swiftness, achieving an average processing time of only 0.10 seconds(s), which has proven itself as an efficient toolkit for users. In certain instances, AutoLoop outperformed competing methods and exhibited satisfactory performance in predicting different dominant loop conformations when proteins bind to different ligands. This capability is crucial for elucidating the dynamics of loop conformation, which plays a significant role in protein function. Overall, AutoLoop has established itself as the leading method in loop prediction, demonstrating unparalleled performance across multiple scenarios.

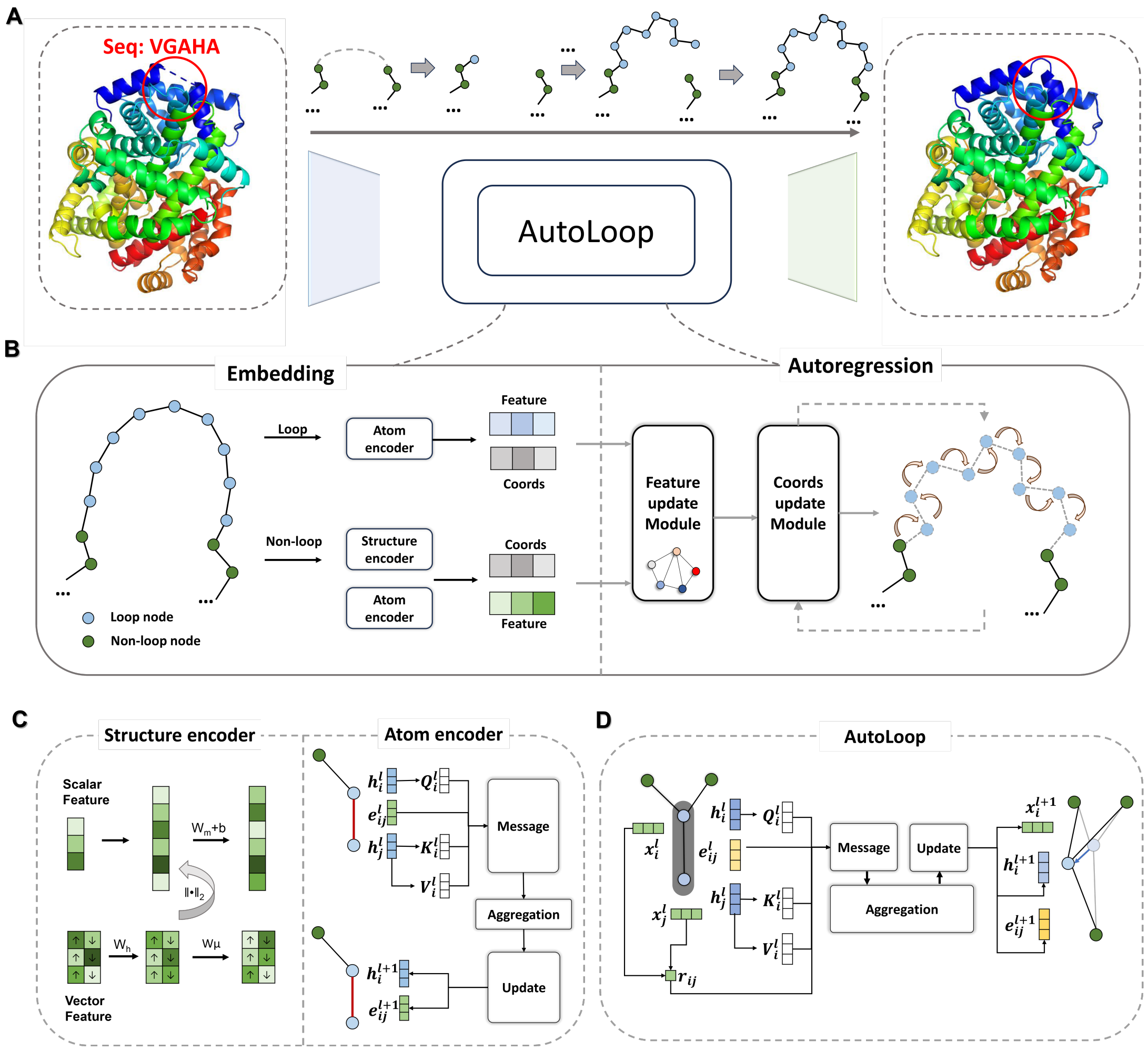

**Figure 1.** Overview of Autoloop's Bidirectional Training Strategy for Loop Conformation Prediction (A) Loop prediction process, a protein with a missing loop and the corresponding loop sequence are inputted into AutoLoop, resulting in the predicted loop conformation. (B) The pipeline of AutoLoop, includes two stages, namely embedding extraction and autoregression. (C) The architecture of structure and atom encoder. (D) The architecture of AutoLoop autoregression layer.

## Results and Discussion

### AutoLoop pipeline

AutoLoop is designed to predict the near-native conformation of protein loops with high precision and efficiency. As illustrated in Figure 1B, AutoLoop unfolds in two primary stages: embedding and autoregression procedure. Initially, in the embedding stage, two types of encoders are utilized: a structure encoder and an atom encoder. The structure encoder processes only the non-loop amino acids, which have known coordinates, while the atom encoder handles the loop regions and adjacent non-loop areas without coordinate information. In the subsequent autoregression phase, specific modules update features and coordinates to generate precise loop conformations. Figure 1C provides a visual description of both the atom and structure encoders. The structure encoder incorporates Geometric Vector Perceptrons[42] (GVP) to encode both scalar and vector features which aim to process both directional and non-directional information, while the atom encoder leverages Graph Transformer[43, 44] to accurately capture atomic-level details, as detailed in Figure 1D. After obtaining the conformation predicted by AutoLoop, the post-processing module can be applied which includes the side-chain packing and the minimization of energy to refine the structure. Each module involved in the autoregression phase is methodically described in the **Method** section to provide a comprehensive understanding.

**Accurate loop conformation prediction**

In the realm of loop prediction within structural bioinformatics, the key factor to consider is the accuracy of the prediction results, as it significantly impacts their utility in subsequent applications. As for AutoLoop, the post-processing module (side-chain packing and energy minimization) is indispensable for obtaining the all-atom conformation of the loop region. As most tested methods are involved in energy minimization but some are not (e.g., FREAD), we evaluated both the initial predictions (denoted as AutoLoop) and the post-processed conformations (denoted as AutoLoop_p) for comprehensive comparison, using two independent datasets: CASP15 and HOMSTARD. The Root Mean Square Deviation (RMSD) was chosen as the metric for the following comparison. Our analysis, as illustrated in Figures 2,3 and 4, focuses on the CASP15 dataset and distinctly highlights Autoloop's superior performance. For instance, Figure 2A shows that Autoloop and Autoloop_p achieved a median RMSD of 1.12 Å and 1.10 Å, respectively. It is significantly better than the competing tools which show median RMSDs ranging from 1.88 Å to 4.08 Å. Furthermore, Figure 2B provides insights into the success rates of these tools at achieving RMSDs no more than the thresholds of 2Å and 1Å, wherein AutoLoop leads with success rates of 73.23% and 48.92% respectively. After post-processing, the 2Å and 1Å success rates slightly changed to 79.19% and 44.41%, respectively, both outperforming the closest competitor, NGK, with the 2Å and 1Å success rates of 52.37% and 20.50%, respectively. Additionally, Figure 2C illustrates the average RMSD and its standard deviation (Std.) for each method, underscoring AutoLoop's enhanced consistency and reliability in predicting accurate loop conformations compared to its counterparts. To delve deeper, we examined the RMSD of the predicted conformations for individual samples between the tested methods and AutoLoop (Figure S1) or AutoLoop_p (Figure S2). This side-by-side RMSD comparison for each sample illustrates the difference in performance among the methods. Our results indicate that, when compared to other methods on individual samples, AutoLoop demonstrates superior performance in at least 81.05% of the cases. Notably, post-processing leads to a reduction in RMSD in 54.01% of the cases involving AutoLoop-predicted conformations.

We further evaluated AutoLoop's ability to make accurate predictions not only in crystal structure environments but also in model-predicted structure environments. As shown in Figure 2C, various protein structure prediction methods (AF2, AF3, RF, CF) were utilized, and their predicted structures were used as inputs for AutoLoop. The results demonstrate that AutoLoop effectively improves the accuracy of loop structures compared to the original predictions. This suggests that AutoLoop can be employed to refine loops in protein structures predicted by other tools. Furthermore, these findings indicate that the overall accuracy of protein structure predictions can significantly influence loop prediction performance.

To extend our insights, we incorporate the HOMSTARD dataset to evaluate AutoLoop alongside additional tools available only through web servers. These tools, which include DaReUS-Loop[16], Sphinx[9], LoopIng[12], and Galaxy PS2[18], were not part of the CASP15 comparison due to dataset size constraints. Moreover, the previous second-best method NGK was also selected. The results, as detailed in Table 1, indicate that AutoLoop consistently outperforms the other methodologies with RMSD values of 2.16 Å (median), 2.34 Å (average), and 0.68 Å (standard deviation) initially, which further improve to 1.39 Å, 1.47 Å, and 0.60 Å, respectively, after post-processing.

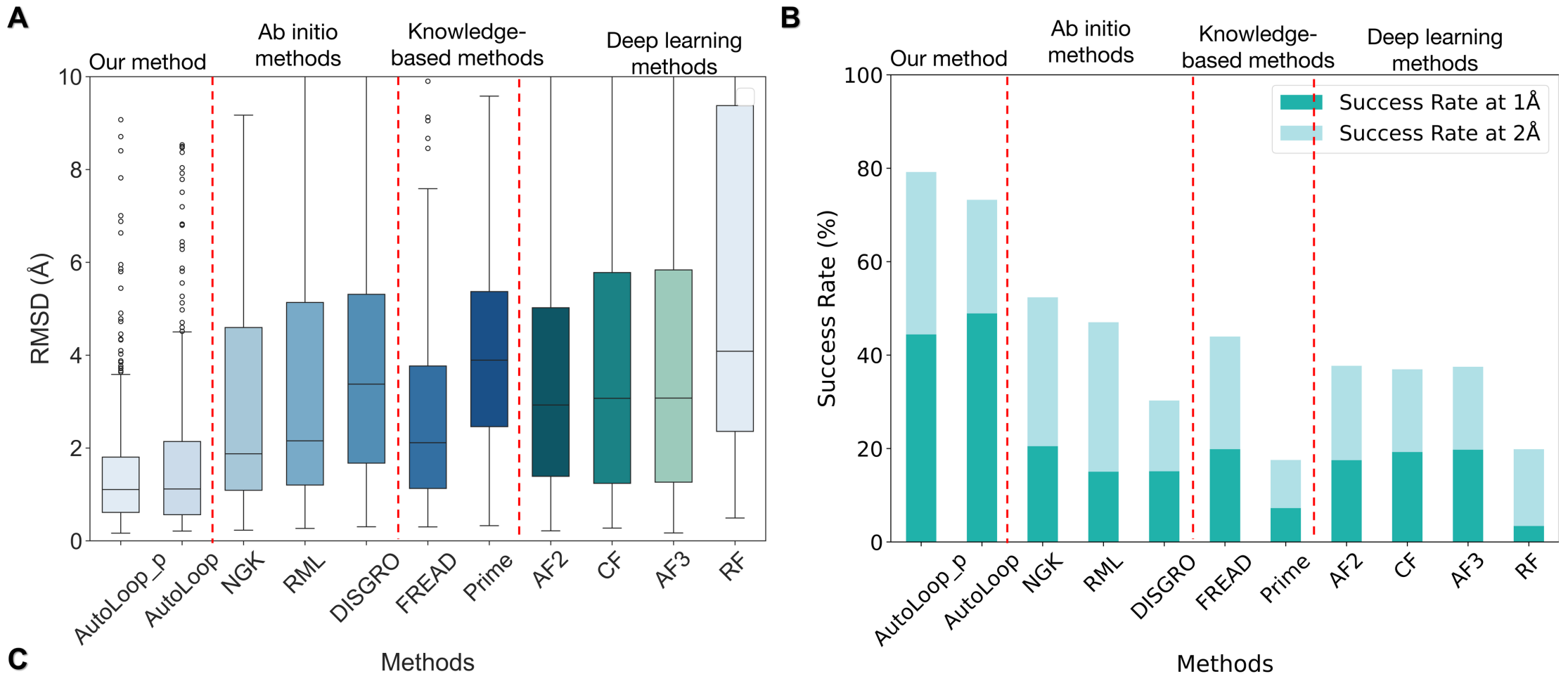

C

| Software | Average (Å) | Std. (Å) |
|---|---|---|
| RF | 9.24 | 13.35 |
| RF (AutoLoop_refined) | 9.05 | 13.17 |
| AF3 | 6.49 | 11.03 |
| AF3 (AutoLoop_refined) | 6.11 | 9.80 |
| CF | 6.34 | 10.87 |
| CF (AutoLoop_refined) | 6.18 | 10.72 |
| AF2 | 5.28 | 7.98 |
| AF2 (AutoLoop_refined) | 4.81 | 6.68 |
| DISGRO | 5.03 | 9.27 |
| Prime | 4.36 | 2.87 |
| RML | 3.75 | 3.91 |
| NGK | 3.40 | 3.64 |
| FREAD | 3.17 | 3.79 |
| AutoLoop | **1.90** | **2.35** |
| AutoLoop_p | **1.53** | **1.51** |

**Figure 2.** Performance of AutoLoop and other tested methods on loop prediction accuracy, with AlphaFold2 denoted as AF2, AlphaFold3 as AF3, ColabFold as CF, RoseTTAFold as RF, for short, and AutoLoop_p indicates AutoLoop with post-processing module. (A) The RMSD distribution of each method. (B) The success rate was calculated in the 1Å and 2Å RMSD threshold of the tested methods. (C) The average RMSD and its standard deviation (Std.) of the tested methods.

**Table 1.** The performance of each method on HOMSTARD, including the Average, Standard deviation (Std.) and Median RMSD values of the tested methods.

| Software | Average (Å) | Std. (Å) | Median (Å) |
|---|---|---|---|
| **LoopIng** | 6.48 | 2.10 | 6.16 |
| **DaReUS-Loop** | 4.42 | 2.08 | 4.34 |
| **Sphinx** | 3.86 | 1.73 | 3.20 |
| **Galaxy PS2** | 3.43 | 1.18 | 3.30 |
| **NGK** | 3.32 | 1.47 | 3.14 |
| **AutoLoop** | **2.34** | **0.68** | **2.16** |
| **AutoLoop_p** | **1.47** | **0.60** | **1.39** |

**Influence of TM-score on prediction accuracy**

The Template Modeling (TM)-score[45] is a widely used metric for assessing the topological similarity between protein structures. Unlike traditional measures such as root-mean-square deviation (RMSD), the TM-score is designed to be independent of protein length, making it particularly useful for comparing proteins of varying sizes. A TM-score value closer to 1 indicates a higher degree of structural similarity. In this study, we utilized US-align[46] to calculate the TM-score between the test set and the training set, which performs structural alignment and computes TM-score. Figure 3 illustrates the trend of average RMSD across different TM-score ranges for the tested methods. The results indicate that, compared to other tested methods, AutoLoop and AutoLoop_p exhibit greater stability across various TM-score ranges and demonstrate superior performance within each range. These findings suggest that the success of AutoLoop is not dependent on sequence or structure similarity but rather on its generalization ability.

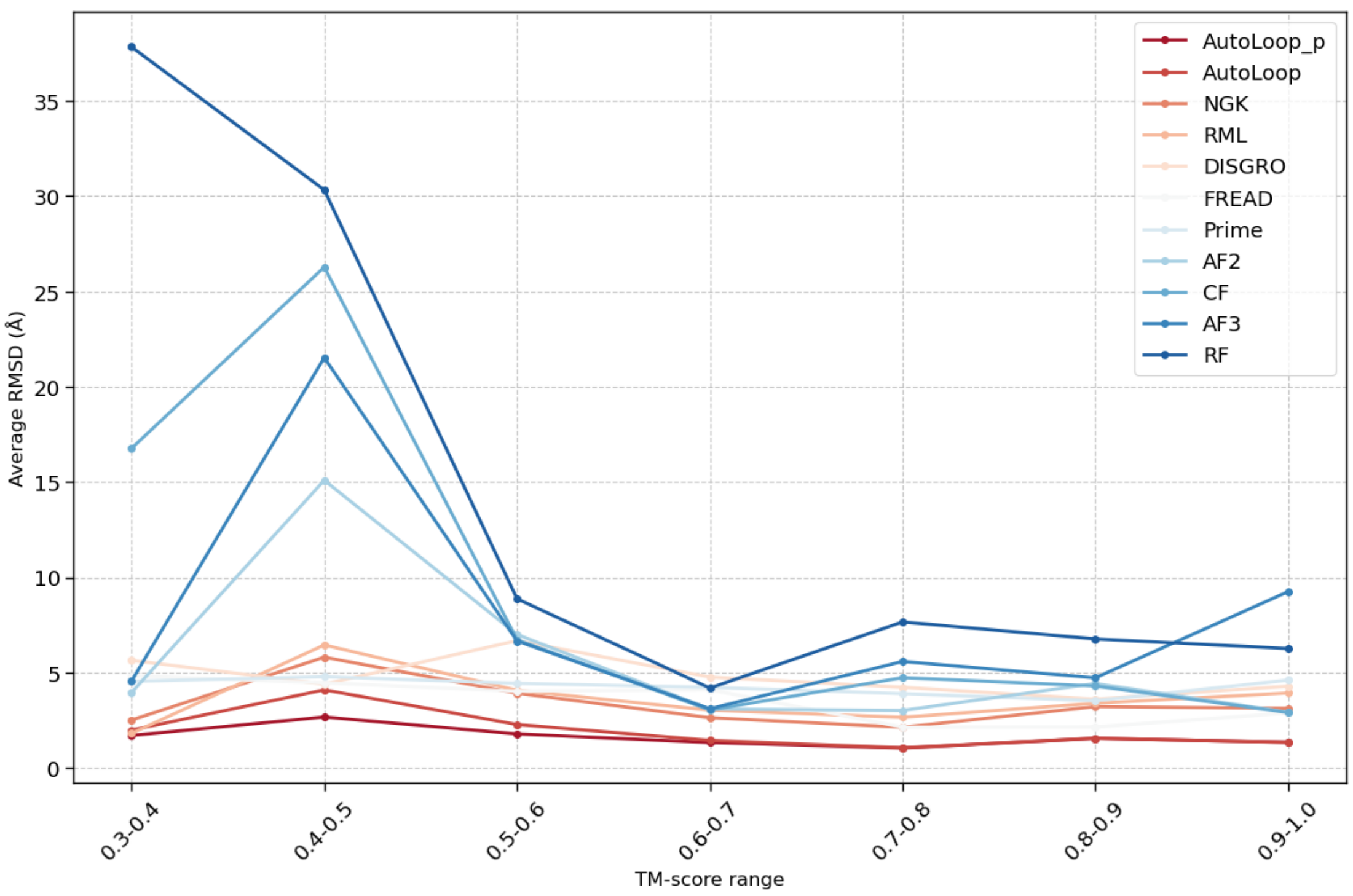

**Figure 3** The performance of the tested methods on different TM-score ranges on CASP15.

**Influence of loop length on prediction accuracy**

The accurate prediction of long loops remains a persistent challenge within the field[6]. To better understand the influence of loop length on prediction accuracy, we concentrated our analysis on the CASP15 test dataset. This focus was necessitated by the limited availability of suitable samples within the HOMSTRAD database. Figure 4 illustrates that AutoLoop outperforms other methods across most loop lengths, irrespective of whether considering the initial or post-processed conformations. Post-processed conformations consistently exhibit superior performance compared to the initial conformations predicted by AutoLoop, across all loop-length samples. Notably, the RMSD values increase slowly with length, highlighting AutoLoop as an effective tool insensitive to loop length. However, it is important to note that the sample size diminishes as loop lengths increase, particularly for loops exceeding 20 amino acid residues, where often only one or a few samples are available. Consequently, this scarcity of data introduces an element of randomness to the results, suggesting that findings should be interpreted with caution.

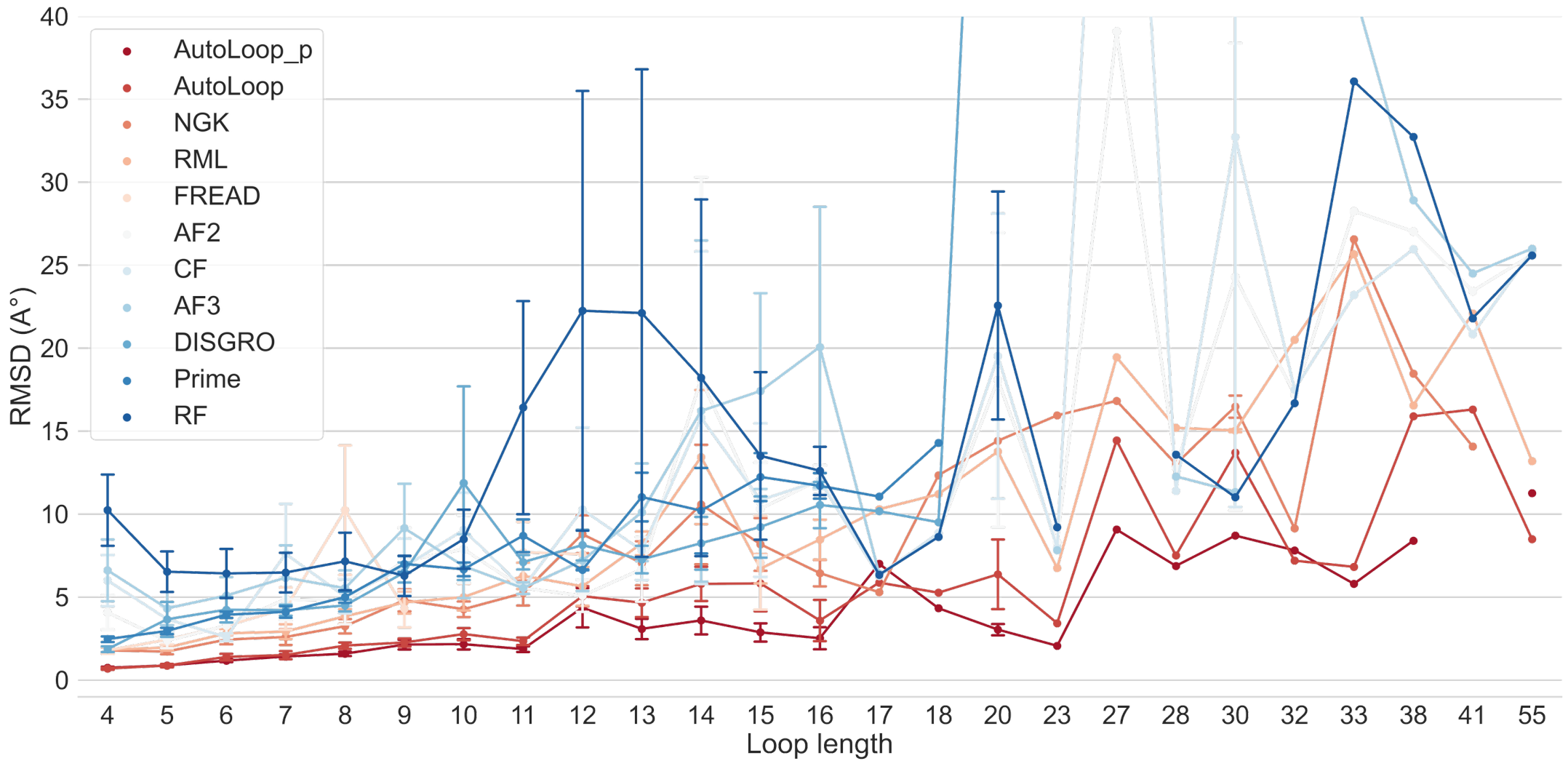

**Figure 4**. The performance of the tested methods on different loop lengths (number of amino acids) on CASP15. (The missing points are due to the corresponding method failing in this loop-length task.)

## Influence of secondary structure on prediction accuracy

According to the definition provided by DSSP[47], loops comprise three secondary structure types: T (Turns), S (Bends), and C (Coils or irregular curls). Turns are relatively short and structured, often facilitating sharp changes in the direction of the polypeptide chain, thus connecting two regular secondary structures like alpha-helices or beta-strands. In contrast, random coils do not adopt a fixed conformation and are characterized by their disorder, while bends represent deviations in the protein backbone that do not fit into standard secondary structure categories. As illustrated in Figure 5, AutoLoop consistently exhibited superior performance across all types of secondary structures, making it a reliable tool for loop prediction regardless of the loop type. Furthermore, the conformations predicted demonstrated the best performance on Turns, while showing varying performance on Coils and Bends. This superior performance on Turns is likely attributable to their more regular and predictable nature compared to Coils and Bends.

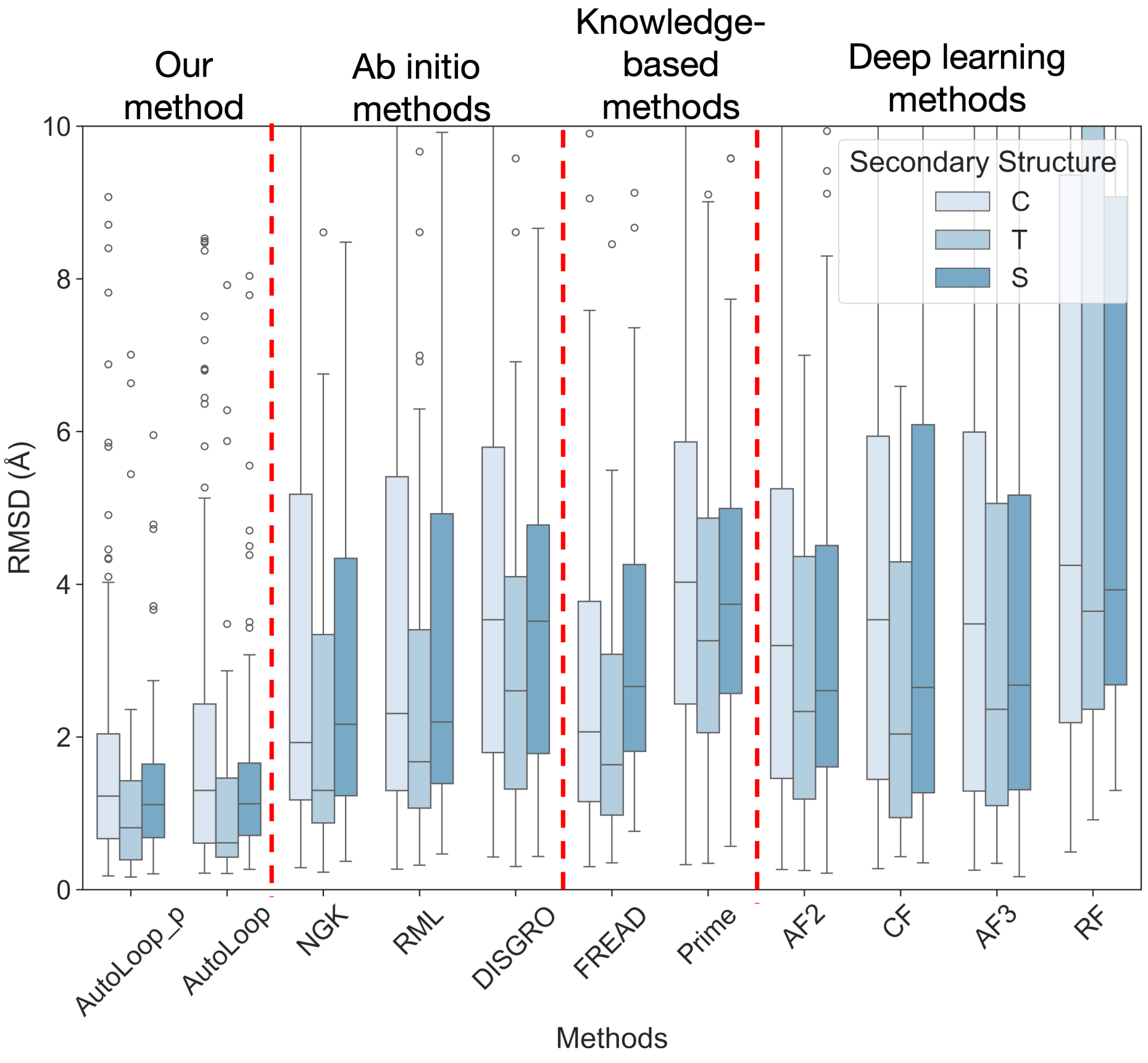

**Figure 5**. The performance of the tested methods on different secondary structures on CASP15. According to the definition from DSSP, C indicates random coils and S indicates bends and T indicates turns.

**Computational efficiency**

Computational efficiency is also crucial in predicting protein loops. This section assesses the time efficiency of various methodologies on CASP15. To accommodate the requirements of users with restricted resources, we utilized AutoLoop on a Tesla a V100S GPU and a single-core Intel(R) Xeon(R) Gold 6240R CPU @ 2.40GHz for loop prediction. While other methods, including FREAD, Prime, $D_ISG_{RO}$, NGK, and RML, were evaluated on an Intel(R) Xeon(R) Gold 6240R CPU @ 2.40GHz, utilizing a single core for each method. Notably, as AlphaFold2, RoseTTAFold and CoabFold are aiming to predict the whole protein structure which requires more substantial computational resources; thus, it was tested using both a Tesla V100S GPU and a 20-core Intel(R) Xeon(R) Gold 6240R CPU @ 2.40GHz. As illustrated in Table 2, AutoLoop demonstrated outstanding time efficiency, with an average processing time of just 0.10s per prediction. Even when incorporating additional computational steps such as adding side chains and relaxing the conformation of the loop with OpenMM[48], the processing time only increased modestly to 0.46s. In contrast, the time consumption in other tested methods ranged significantly, from 21.579s to 1172.342s. To be clear, AlphaFold3 was deployed as a web server, making it challenging to directly calculate the time for prediction. Typically, each AlphaFold3 task takes approximately 120s-240s to completion, with pending time included. Such a dramatic improvement emphasizes the transformative potential of deep learning-based approaches in significantly expediting computational tasks within this domain.

**Table 2**. The average time cost of each method for completing one task.

| Software | Average time (s) |
|---|---|
| **AF2** | 3337.25 |
| **RF** | 1844.74 |
| **NGK** | 978.43 |
| **RML** | 930.80 |
| **CF** | 192.02 |
| ***AF3** | 120-240 |
| **Prime** | 120.66 |
| **FREAD** | 61.43 |
| **D$_{IS}$G$_{RO}$** | 21.58 |
| **AutoLoop_p** | **0.46** |
| **AutoLoop** | **0.10** |

* The AF3 web-server does not provide precise information about the processing duration, and the estimated time consumption in the table includes pending time.

## Case study on accurate loop conformation prediction

Building on the comparisons previously discussed, it is evident that AutoLoop is both accurate and efficient in predicting protein loop conformations. This is further illustrated through visual comparisons of AutoLoop's predictions with those from other methodologies, as juxtaposed with actual (ground-truth) conformations. Figure 6 presents various loop length cases from the CASP15 dataset, specifically with loop lengths of six (T1123-D1), seven (T1188-D1), nine (T1139-D1), twenty (T1157s1-D1), and thirty-three (T1137s1-D2) residues. The visual comparisons show that the conformations predicted by AutoLoop (green) and post-processed AutoLoop (blue) more closely align with the ground-truth conformations (red) and consistently outperform those predicted by NGK (wheat) and AlphaFold2 (purple) across different loop lengths.

Beyond this, another widely discussed question is the dynamics of loop conformations in proteins, which may change when the protein binds to different ligands, causing the loop region to exhibit different structures. To illustrate this, we focus on the RAF proto-oncogene serine/threonine-protein kinase, commonly known as RAF kinase. This kinase is a crucial component of the RAS-RAF-MEK-ERK signaling pathway, which plays a significant role in regulating cell division, differentiation, and survival[49]. As a critical target in cancer research, RAF kinase has driven the development of inhibitors aimed at disrupting its abnormal signaling in cancer cells. In this study, we selected PDB ID 6VJJ (a complex of RAF kinase and GTPase KRas) and 3KUC (a complex of RAF kinase and Ras-related protein Rap-1A) to investigate whether AutoLoop can incorporate environmental elements to predict loop conformations accurately. As illustrated in Figure 7A, the loops in these two proteins exhibit significantly different conformations, with an RMSD of 4.69 Å. In the 3UKC RAF protein (chain B), mutations A85K and N71R introduce changes not present in the 6VVJ structure. The 3KUC RAF protein (chain B) contains A85K and N71R mutations not present in the 6VVJ structure. Through structural analysis, we observed that these differences correlate with distinct interaction patterns. In 3KUC, residues in the vicinity of positions 85 and 71 form alternative hydrogen bonding networks compared to 6VJJ. The residue at position 74N in 3KUC adopts a different position that facilitates interaction with 117N. While these structural differences correlate with the observed loop conformation changes, it's important to note that multiple factors could contribute to these variations, including crystallization conditions, crystal packing effects, and inherent protein flexibility. The observed differences highlight the challenge in loop modeling and demonstrate why considering the broader structural context, as implemented in our fully connected graph approach, may provide valuable information for predicting loop conformations under various conditions (Figure S3; further details are provided in the Supplementary Information).

Figures 7B and 7C demonstrate that AutoLoop can recognize different environment factors and predict near-native conformations, achieving RMSDs of just 1.51 Å and 0.87 Å, respectively. This performance demonstrates AutoLoop's effectiveness in incorporating environmental influences to generate more precise predictions that closely reflect actual structural conformations.

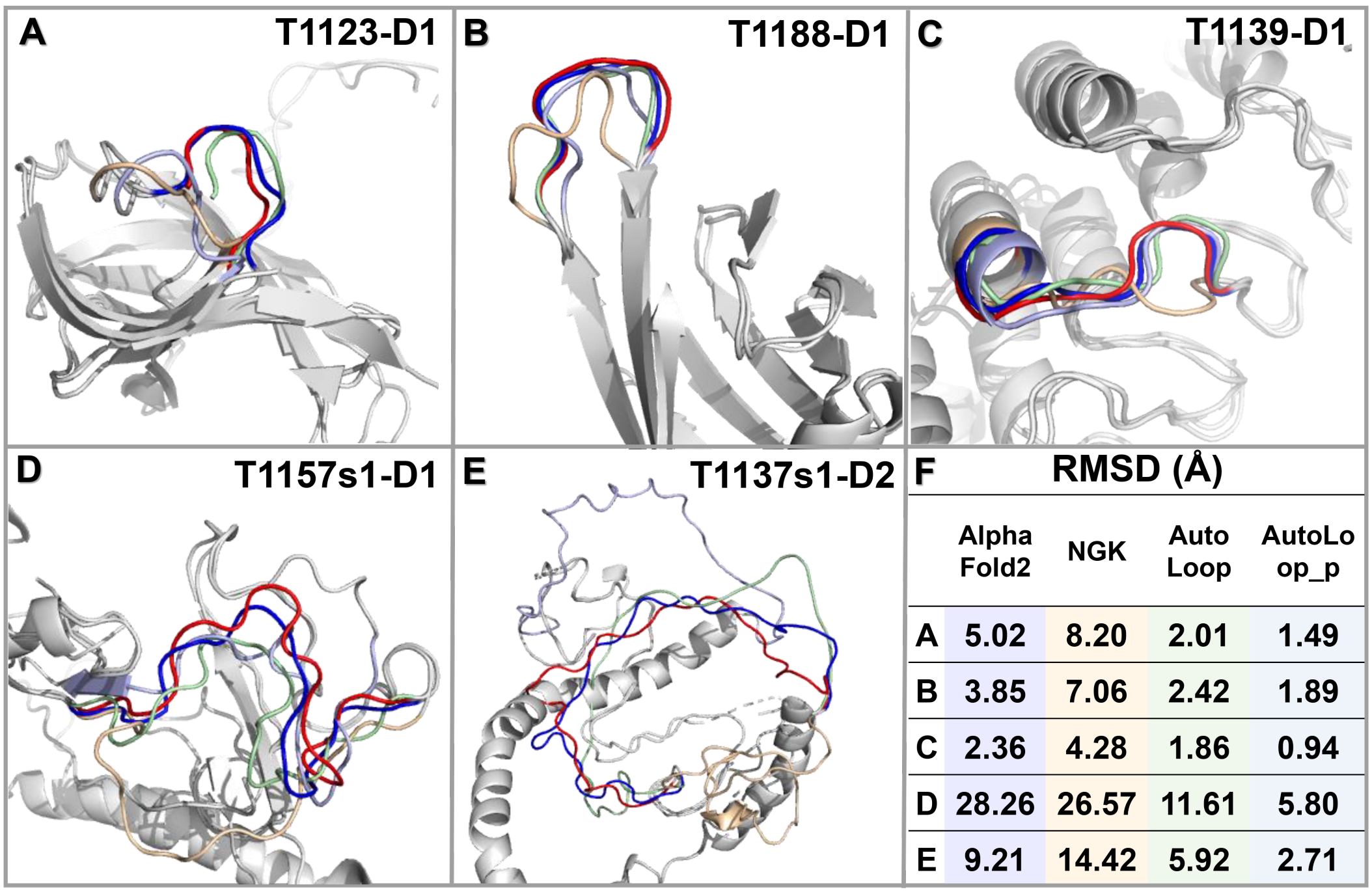

| | Alpha Fold2 | NGK | Auto Loop | AutoLo op_p |
|---|---|---|---|---|
| A | 5.02 | 8.20 | 2.01 | 1.49 |
| B | 3.85 | 7.06 | 2.42 | 1.89 |
| C | 2.36 | 4.28 | 1.86 | 0.94 |
| D | 28.26 | 26.57 | 11.61 | 5.80 |
| E | 9.21 | 14.42 | 5.92 | 2.71 |

**Figure 6**. Examples of different length loop prediction results: Ground-truth (red), AutoLoop_post (blue), AutoLoop (green), NGK (wheat) and AlphaFold2 (purple). (A) CASP15 ID: T1123-D1 (loop length: 6 residues). (B) CASP15 ID: T1181-D1 (loop length: 7 residues). (C) CASP15 ID: T1139-D1 (loop length: 9 residues). (D) CASP15 ID: T1157s1-D1 (loop length: 20 residues). (E) CASP15 ID: T1137s1-D2 (loop length: 33 residues). (F) RMSD values of the tested methods compared with the ground-truth loop conformations.

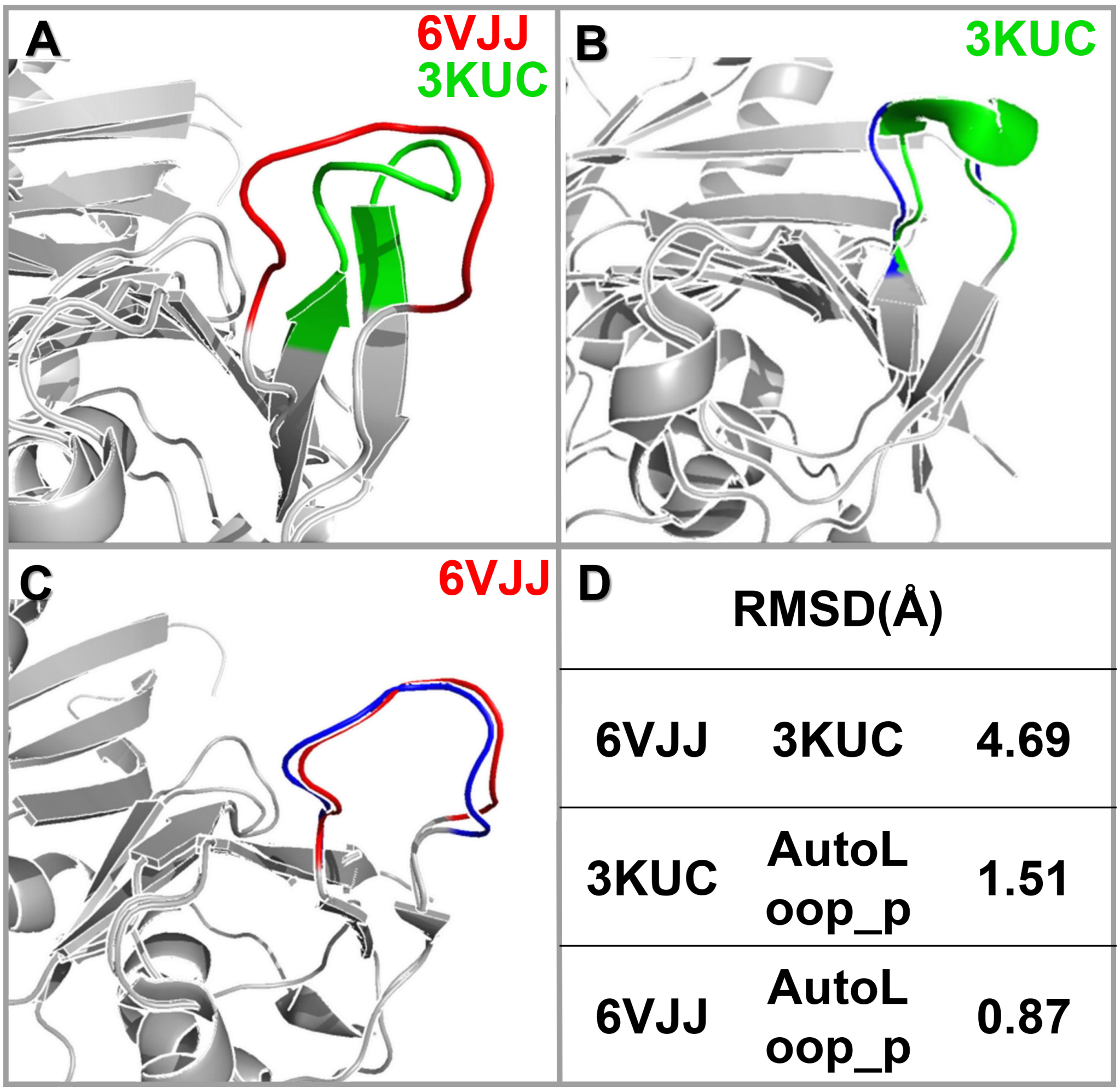

**Figure 7**. Different dominant loop conformation prediction: green (PDB ID: 3KUC), red (PDB ID: 6VJJ), blue (AutoLoop_post): (A) 3KUC vs 6VJJ, (B) 3KUC vs AutoLoop_post, and (C) 6VJJ vs AutoLoop_post. (D) The RMSD values between those conformations.

## Conclusion

Loop prediction is one of the most difficult questions in protein structure modeling. Thus, we developed AutoLoop, a novel DL method that utilizes autoregression for making accurate loop predictions. This method is further enhanced by a bi-directional training strategy, designed to improve both robustness and performance. To evaluate AutoLoop, we conducted tests using two widely recognized datasets: CASP15 and HOMSTRAD, alongside comparisons with thirteen established methods. The results indicate that AutoLoop achieves 1.12Å median RMSD, 1.90 Å average RMSD, and 2 Å and 1 Å success rates of 73.23% and 48.92% on the CASP15 dataset. The addition of a post-processing module slightly enhances its performance, decreasing the median and average RMSD to 1.10 Å, 1.53 Å, respectively. The results indicate that AutoLoop represents an effective deep-learning approach for predicting loop regions, with its accuracy primarily attributed to the novel algorithm rather than to energy minimization techniques. Moreover, AutoLoop shows substantial superiority on the HOMSTRAD dataset, with median and average RMSD improvements of at least 40.11% and 40.06% before post-processing, and 41.89% and 51.73% afterward. AutoLoop is also capable of accurately predicting loops with different lengths and different structure similarity thresholds, which is verified on CASP15 that AutoLoop outperforms other tested methods on most lengths of loop samples and all TM-score ranges. The RMSD values increase slowly with the increase of the loop length and decrease of structure similarity, indicating the insensitivity of the model to loop length and structure similarity. AutoLoop also outperforms other methods against all types of secondary structures. Notably, AutoLoop completes tasks in an average of 0.10s, which demonstrates the outstanding efficiency of AutoLoop. Additionally, AutoLoop shows the potential to predict different dominant conformations influenced by environmental variables, like differing ligands. This adaptability is crucial for understanding protein functionality under various conditions. Beyond its precise predictive capabilities, AutoLoop holds significant potential for broader applications in molecular dynamicsand drug design by potentially generating near-native loop conformations.

Despite the considerable advancements achieved, AutoLoop continues to exhibit certain limitations. The model currently only provides a single conformation. To overcome this limitation, future work could involve developing a probabilistic model that generates multiple conformations. Such advancements would broaden AutoLoop's utility, paving the way for more dynamic and versatile applications in protein modeling. Future research will therefore focus on the creation of this probabilistic model.

## Methods

### Dataset

The dataset used in this study was divided into training data and testing data. The training dataset was derived from PDB in 2022.08 which contains more than 210,000 experimental-determined structures and filtered by PISCES[50] according to the following standard: solved by X-ray crystallography, sequence identity ≤90%, resolution <3.0 Å, R-factor≤0.25. Loop region was defined by DSSP[51] and no less than 4 amino acids. All those data were collected in 2022.8, and the training and validation datasets were created using random stratified sampling according to the loop length in a 9:1 ratio. The test data was derived from CASP15, HOMSTARD The CASP15 dataset was derived from the CASP15 monomer dataset and was further filtered by the threshold of chain sequence identity of no more than 40% between itself and the training dataset which contains 319 loop structures from 35 proteins (the unmatched structure and sequence sample were removed) and HOMSTARD provides 18 loop structures from 16 proteins, ensuring no overlap with the training data.

All the protein data was downloaded from PDB and CASP datasets. Samples containing non-standard amino acids within the loop regions were subsequently excluded through a filtering process. Before training, the protein pocket was selected to reduce the computational cost. Specifically, amino acids within a 12 Å radius of the loop regions were chosen as pockets, taking into account both short-range and long-range interactions.

**AutoLoop architecture**

AutoLoop employs an end-to-end DL approach to predict protein loop conformation, leveraging a bidirectional training strategy to improve robustness. As shown in Figure 1. AutoLoop has four stages in total (1) graph generation module, (2) embedding module, (3) conformation prediction (autoregression) module (4) post-processing module. The autoregression module consists of 8 layers to apply features and coordinate updates. The details about those three modules are illustrated below.

**Graph generation module**

This methodology employs a Graph Neural Network (GNN) framework to represent proteins as undirected graphs, wherein atoms are depicted as nodes and covalent bonds form the edges. Achieving precise predictions of loop conformations, especially with a focus on non-bonded interactions, necessitates the construction of a comprehensive network that fully connects all nodes, inclusive of those within non-loop and loop regions. This approach poses significant computational challenges, arising from the substantial number of atoms that proteins typically comprise. To mitigate this complexity, residue graphs were formulated for non-loop nodes by conceptualizing residues as nodes. Edges between these nodes were defined using the K-Nearest Neighbor (KNN, K=30) algorithm, based on the spatial proximity of the residues' alpha carbon (CA) coordinates. Such a strategy notably simplifies the graph's complexity by curtailing the total number of nodes and edges, while also integrating the geometric nuances of each residue.

Furthermore, the method integrates atom-level information from the protein graph with residue-level attributes derived from residue graphs for non-loop nodes, thereby equipping the model to discern interactions with both coarse-grained and fine-grained specificity. This amalgamated configuration is especially advantageous for the detection of long-range inter-residue interactions, irrespective of their location within the loop regions or beyond. Consequently, this enhances the model's capability to accurately forecast loop conformations.

**Embedding module**

The embedding module is primarily composed of two layers: the Graph Transformer (GT) and the Geometric Vector Perceptron (GVP). These layers are designed to extract features from both atoms and residues. It is anticipated that this multi-scale modeling approach will reduce computational costs and improve accuracy

**Graph Transformer (GT)**

GT was implemented for intramolecular interaction learning (Figure 1C) in terms of protein graphs, including both non-loop and loop atoms. The node feature $n_i \in R^{d_h \times 1}$ for $i$th node and edge feature $e_{ij} \in R^{d_e \times 1}$ for the edge between node $i$ and node $j$ are first initialized to $h^0_{p_i}$ and $e^0_{p_ij}$ with d-dimension by two linear layers, respectively:

$$n_i^0 = W_h^0 n_i + b_h^0$$
$$e_{ij}^0 = W_e^0 e_{ij} + b_e^0 \tag{1}$$

where $W_h^0 \in R^{d \times d_h}$, $W_e^0 \in R^{d \times d_e}$ and $b_h^0$, $b_e^0 \in R^d$. The initialized node and edge embedding are then input to the graph transformer layer stacking six times to output the final embeddings of nodes and edges. The $l$th graph transformer layer updates node edge embeddings using message passing and a modified multi-head self-attention (MHA) mechanism, as shown in the following equations:

$$q_i^{k,l} = W_Q^{k,l} Norm(n_i^l) \tag{2}$$

$$k_j^{k,l} = W_K^{k,l} Norm(n_j^l) \tag{3}$$

$$v_j^{k,l} = W_V^{k,l} Norm(n_j^l) \tag{4}$$

$$e_{ij}^{k,l} = W_E^{k,l} Norm(e_{ij}^l) \tag{5}$$

$$w_{ij}^{k,l} = Softmax_{j \epsilon N(i)}\left(\left(\frac{q_i^{k,l} \cdot k_j^{k,l}}{\sqrt{d_k}}\right) \cdot e_{ij}^{k,l}\right) \tag{6}$$

$$n_i^{l+1} = h_i^l + W_{h0}^l Dropout\left(Concat_{k \epsilon 1,\ldots,H}\left(Aggregation_Sum_{j \epsilon N(i)}\left(w_{ij}^{k,l} v_j^{k,l}\right)\right)\right) \tag{7}$$

$$\hat{e}_{ij}^{l+1} = e_{ij}^l + W_{e0}^l Dropout\left(Concat_{k \epsilon 1,\ldots,H}\left(w_{ij}^{k,l}\right)\right) \tag{8}$$

$$n_i^{l+1} = \hat{n}_i^{l+1} + W_{h2}^l Dropout\left(SiLU\left(W_{h1}^l Norm\left(\hat{n}_i^{l+1}\right)\right)\right) \tag{9}$$

$$e_{ij}^{l+1} = \hat{e}_{ij}^{l+1} + W_{e2}^l Dropout\left(SiLU\left(W_{e1}^l Norm\left(\hat{e}_{ij}^{l+1}\right)\right)\right) \tag{10}$$

where $W_Q^{k,l}$, $W_K^{k,l}$, $W_V^{k,l}$, $W_E^{k,l} \epsilon R^{d_k \times d}$, $W_{h0}^l$, $W_{e0}^l \epsilon R^{d \times d}$, $W_{h1}^l$, $W_{e1}^{k,l} \epsilon R^{2d \times d}$, and $W_{h2}^l$, $W_{e2}^l \epsilon R^{d \times 2d}$ are learnable parameters from linear layers; $k \epsilon 1, \ldots, H$ denotes the number of attention heads; $d_k$ is the dimension of each head, which equals $d$ divided by $H$; $j \epsilon N(i)$ represents the neighboring nodes of node $i$; $Norm$ denotes batch normalization; $Concat$ denotes the concatenation operation; $Dropout$ denotes the dropout operation; $SiLU$ represents a type of activation functions; $Aggregation_Sum_{j \epsilon N(i)}$ represents aggregating the messages on the edges connecting node $i$ and its neighboring nodes $j$ by summation; and $Softmax_{j \epsilon N(i)}$ denotes the SoftMax operation on neighboring nodes $j$.

**Geometric Vector Perceptrons (GVP)**

GVP is implemented in AutoLoop as the residue encoder to update the non-loop embeddings based on the topology connections and geometric features inside and between resides. The basic block of GVP is the gvp layer that receives both scalar features $f_s \in R^d$ and vector features $f_v \in R^{d\times3}$. The forward process of $l$th layer is as follows:

$$f_{v1}^l = W_{v0}^l f_v^l \quad (11)$$

$$f_{v2}^l = W_{v1}^l f_{v1}^l \quad (12)$$

$$S_{v1}^l = \|f_{v1}^l\|_2 (row\ wise) \quad (13)$$

$$S_{v2}^l = \|f_{v2}^l\|_2 (row\ wise) \quad (14)$$

$$f_{sv}^l = Concat(f_s^l, S_{v1}^l) \quad (15)$$

$$\hat{f}_s^l = W_{sv}^l f_{sv}^l + b_{sv}^l \quad (16)$$

$$f_s^{l+1} = \sigma_s(\hat{f}_s^l) \quad (17)$$

$$f_v^{l+1} = \sigma_v(S_{v2}^l) \odot f_{v2}^l\ (row\ wise\ multiplication) \quad (18)$$

where $W_{v0}^l \epsilon R^{d_{v1}\times d_{v0}}$, $W_{v1}^l \epsilon R^{d_{v2}\times d_{v1}}$ and $W_{sv}^l \epsilon R^{d_{s1}\times(d_{s0}+d_{v1})}$ are learnable parameters; $f_{v1}^l \epsilon R^{d_{v1}\times3}$, $f_{v2}^l \epsilon R^{d_{v2}\times3}$, $S_{v1}^l \epsilon R^{d_{v1}}$, $S_{v2}^l \epsilon R^{d_{v2}}$, $f_{sv}^l \epsilon R^{d_{s0}+d_{v1}}$, $b_{sv}^l \epsilon R^{d_{s1}}$, $\hat{f}_s^l \epsilon R^{d_{s1}}$, $f_v^{l+1} \epsilon R^{d_{v2}\times3}$ and $f_s^{l+1} \epsilon R^{d_{s1}}$ are the results of equations; $\sigma_s$ and $\sigma_v$ represent the activation functions. Before being input to gvp layers, the sequence information is embedded by the Embedding layer and concatenated with other scalar node features:

$$h_{seq} = Embedding(Sequence) \quad (19)$$

$$h_s = Concat(h_{s0}, h_{seq}) \quad (20)$$

where the dimension of the word table of the Embedding layer is $(d_{seq}, d_{seq})$ and $h_{s0} \epsilon R^{d_{hs0}}$. Then, the node features and edge features are input to the initialization block consisting of a LayerNorm and a gvp layer without activation functions, respectively:

$$(h_{s1}, h_{v1}) = gvp(LayerNorm(h_s, h_v)) \quad (21)$$

$$(e_{s1}, e_{v1}) = gvp(LayerNorm(e_s, e_v)) \quad (22)$$

where $h_s \epsilon R^{d_{seq}+d_{hs0}}$, $h_v \epsilon R^{d_{hv0}\times3}$, $h_{s1} \epsilon R^{d_{hs1}}$, $h_{v1} \epsilon R^{d_{hv1}\times3}$, $e_s \epsilon R^{d_{es0}}$, $e_v \epsilon R^{d_{ev0}\times3}$, $e_{s1} \epsilon R^{d_{es1}}$ and $e_{v1} \epsilon R^{d_{ev1}\times3}$. After initialization, the node and edge features are input to the GVPConv layer stacking two times involving gvp layers in message passing. The equations of the GVPConv layer are listed as follows:

$$m_{s_ij}^{l} = Concat\left(h_{s_i}^{l}, e_{s_ij}^{l}, h_{s_j}^{l}\right) \quad (23)$$

$$m_{v_ij}^{l} = Concat\left(h_{v_i}^{l}, e_{v_ij}^{l}, h_{v_j}^{l}\right) \quad (24)$$

$$\left(m_{s_ij_1}^{l}, m_{v_ij_1}^{l}\right) = gvp\left(m_{s_ij}^{l}, m_{v_ij}^{l}\right) \quad (25)$$

$$\left(m_{s_ij_2}^{l}, m_{v_ij_2}^{l}\right) = gvp\left(m_{s_ij_1}^{l}, m_{v_ij_1}^{l}\right) \quad (26)$$

$$\left(m_{s_ij_3}^{l}, m_{v_ij_3}^{l}\right) = gvp\left(m_{s_ij_2}^{l}, m_{v_ij_2}^{l}\right) \quad (27)$$

$$\left(\hat{h}_{s_j}^{l}, \hat{h}_{v_j}^{l}\right) = Aggregation_Mean_{i\epsilon N(j)}\left(m_{s_ij_3}^{l}, m_{v_ij_3}^{l}\right) \quad (28)$$

$$\left(\hat{f}_{s_j_0}^{l}, \hat{f}_{v_j_0}^{l}\right) = LayerNorm\left(h_{s_j}^{l} + Dropout\left(\hat{h}_{s_j}^{l}\right), h_{v_j}^{l} + Dropout\left(\hat{h}_{v_j}^{l}\right)\right) \quad (29)$$

$$\left(\hat{f}_{s_j_1}^{l}, \hat{f}_{v_j_1}^{l}\right) = gvp\left(\hat{f}_{s_j_0}^{l}, \hat{f}_{v_j_0}^{l}\right) \quad (30)$$

$$\left(\hat{f}_{s_j_2}^{l}, \hat{f}_{v_j_2}^{l}\right) = gvp\left(\hat{f}_{s_j_1}^{l}, \hat{f}_{v_j_1}^{l}\right) \quad (31)$$

$$\left(h_{s_i}^{l+1}, h_{v_j}^{l+1}\right) = LayerNorm\left(h_{s_j}^{l} + Dropout\left(\hat{f}_{s_j_2}^{l}\right), h_{v_j}^{l} + Dropout\left(\hat{f}_{v_j_2}^{l}\right)\right) \quad (32)$$

where equations (25), (26) and (30) implement the activation functions of ReLU and Sigmoid for scalar features and vector features, respectively, and the other gvp layers use no activation functions; $m_{s_ij}^{l} \epsilon R^{2d_{hs1}+d_{es1}}$, $m_{v_ij}^{l} \epsilon R^{(2d_{hv1}+d_{ev1})\times 3}$, $m_{s_ij_1}^{l}$, $m_{s_ij_2}^{l}$, $m_{s_ij_3}^{l}$, $\hat{h}_{s_j}^{l}$, $\hat{f}_{s_j_0}^{l}$, $\hat{f}_{s_j_2}^{l}$ and $h_{s_i}^{l+1}\ \epsilon R^{d_{hs1}}$, $m_{v_ij_1}^{l}$, $m_{v_ij_2}^{l}$, $m_{v_ij_3}^{l}$, $\hat{h}_{v_j}^{l}$, $\hat{f}_{v_j_0}^{l}$, $\hat{f}_{v_j_2}^{l}$ and $h_{v_j}^{l+1} \epsilon R^{d_{hv1}\times 3}$, $\hat{f}_{s_j_1}^{l} \epsilon R^{4d_{hs1}}$, and $\hat{f}_{v_j_1}^{l} \epsilon R^{(2d_{hv1})\times 3}$; and $Aggregation_Mean_{i\epsilon N(j)}$ represents averaging the messages on the edges connecting node $j$ and its neighboring nodes $i$.

## Autoregression

### Autoregression module

The autoregression module was employed to sequentially predict the positions of atoms within the loop, based on the interactions exerted between pairs of nodes. Initially, we characterized the primary focal atoms as those belonging to the backbone of non-loop residues that are covalently bonded to loop residues; specifically, the nitrogen atom oriented from the C-terminus to the N-terminus and, conversely, the carbon atom from the N-terminus to the C-terminus. After this, we executed multiple autoregression iterations to accurately forecast the location of the backbone atoms (i.e., N, CA, C, O) associated with loop residues, following the predetermined directional sequence. To elaborate, the positions of atoms next to the focal atoms were stochastically initiated around these focal atoms. Consequently, unit vectors designating the direction of motion were predicted, combined with the bond lengths, and subsequently incorporated into the atoms' coordinates to predict their positions. These updated atoms were then designated as the new focal atoms, thereby allowing for the iterative progression of positional adjustments to the subsequent atoms.

In AutoLoop, we utilized E(n) Equivariant Graph Neural Network (EGNN) to implement autoregression, a classical architecture widely used in dynamic systems. Additionally, this study incorporates self-attention into the message-passing mechanism of the EGNN to further improve its performance.

Regarding the initialization of scalar node embeddings $n_0$ of non-loops and loops updated by GVP and GT, respectively, are first initialized through graph normalization techniques. Concurrently, the edge features $e_0$ are initialized by a linear layer:

$$n_1 = GraphNorm(n_0) \quad (33)$$

$$e_1 = W_{e_init} e_0 + b_{e_init} \quad (34)$$

where $n_0$ , $n_1$ and $e_1 \epsilon R^{d_h}$ , $e_0 \epsilon R^{d_e}$ ; $W_{e_init} \epsilon R^{d_h \times d_e}$ and $b_{e_init} \epsilon R^{d_h}$ are learnable parameters in a linear layer. This structured approach ensures a consistent and effective foundation for subsequent model training and conformation prediction.

The basic block used for updating loop coordinates consists of 8 EGNN layers stacked sequentially. Notably, the first 7 layers do not update nodes' positions and only the positions of backbone heavy atoms in the loop region were updated in the 8th EGNN layer. For the loop backbone atoms that are not predicted yet, the position information will be masked in case of data leakage. The message passing process of $l$th EGNN layer is shown as follows:

$$\left(q_i^{k,l}\right)_{k \epsilon 1,\dots,H} = W_Q^l h_{i_1}^l + b_Q^l \quad (35)$$

$$\left(k_j^{k,l}\right)_{k \epsilon 1,\dots,H} = W_K^l h_{j_1}^l + b_K^l \quad (36)$$

$$\left(v_j^{k,l}\right)_{k\epsilon 1,\dots,H} = W_V^l h_{j_1}^l + b_V^l \tag{37}$$

$$e_{ij}^l = Concat\left(e_{ij_1}^l, \left\|x_i^l - x_j^l\right\|_2\right) \tag{38}$$

$$m_{ij}^l = W_{m2}^l\left(LeakyReLU\left(Dropout\left(W_{m1}^l e_{ij}^l + b_{m1}^l\right)\right)\right) + b_{m2}^l \tag{39}$$

$$\left(k_{ij}^{k,l}\right)_{k\epsilon 1,\dots,H} = Concat_{k\epsilon 1,\dots,H}\left(k_j^{k,l}\right) \odot m_{ij}^l \tag{40}$$

$$w_{ij}^{k,l} = \left(q_i^{k,l} \odot k_{ij}^{k,l}\right)/\sqrt{d_k} \tag{41}$$

$$\alpha_{ij}^{k,l} = Softmax_{j\epsilon N(i)}\left(\left\|w_{ij}^{k,l}\right\|_2\right) \tag{42}$$

$$\hat{h}_{i_1}^l = Dropout\left(W_h^l\left(Concat_{k\epsilon 1,\dots,H}\left(Aggregation_Sum_{j\epsilon N(i)}\left(w_{ij}^{k,l} \odot v_j^{k,l}\right)\right)\right) + b_h^l\right) \tag{43}$$

$$h_{i_1}^{l+1} = Gate_Block\left(h_{i_1}^l, \hat{h}_{i_1}^l\right) \tag{44}$$

$$e_{ij_1}^{l+1} = W_e^l Concat_{k\epsilon 1,\dots,H}\left(\alpha_{ij}^{k,l}\right) + b_e^l \tag{45}$$

$$x_i^{l+1} = Coords_Update_Block\left(\left(w_{ij}^{k,l}\right)_{\substack{k\epsilon 1,\dots,H, \\ j\epsilon N(i)}}, x_i^l, \left(x_j^l\right)_{j\epsilon N(i)}\right) \tag{46}$$

where $\alpha_{ij}^{k,l} \epsilon R^1$; $h_{i_1}^l$, $h_{j_1}^l$, $e_{ij_1}^l$, $e_{ij}^l$, $m_{ij}^l$, $\hat{h}_{i_1}^l$, $h_{i_1}^{l+1}$, and $e_{ij_1}^{l+1} \epsilon R^{d_h}$; $q_i^{k,l}$, $k_j^{k,l}$, $k_{ij}^{k,l}$, $v_j^{k,l}$, $w_{ij}^{k,l} \epsilon R^{d_k}$; $W_Q^l$, $W_K^l$, $W_V^l$, $W_{m2}^l$, $W_h^l$, and $W_e^l \epsilon R^{d_h \times d_h}$, $W_{m1}^l \epsilon R^{d_h \times (d_h+1)}$, $b_Q^l$, $b_K^l$, $b_V^l$, $b_{m1}^l$, $b_{m2}^l$, $b_h^l$, and $b_e^l \epsilon R^{d_h}$ are learnable parameters from linear layers; $k\epsilon 1, \dots, H$ denotes the number of attention heads; $d_k$ is the dimension of each head, which equals $d_h$ divided by $H$; $j\epsilon N(i)$ represents the neighboring nodes of node $i$; $Concat$ denotes the concatenation operation; $Dropout$ denotes the dropout operation; $LeakyReLU$ is a type of activation functions; $Aggregation_Sum_{j\epsilon N(i)}$ represents summing the messages on the edges connecting node $i$ and its neighboring nodes $j$; and $Softmax_{j\epsilon N(i)}$ denotes the SoftMax operation on neighboring nodes $j$. It should be noted that formula (46) was only performed in the 8th layer.

The updating of the coordinates of the loop nodes procedure is as follows:

$$\Delta\overrightarrow{x_{ij}^l} = x_i^l - x_j^l \tag{47}$$

$$\Delta\overrightarrow{x_{ij}^l} = \Delta\overrightarrow{x_{ij}^l}/\left\|\Delta\overrightarrow{x_{ij}^l}\right\|_2 \tag{48}$$

$$\Delta x_{ij}^{k,l} = \Delta\overrightarrow{x_{ij}^l} \cdot \left(W_{x2}^l\left(LeakeyReLU\left(Dropout\left(W_{x1}^l w_{ij}^{k,l} + b_{x1}^l\right)\right)\right) + b_{x2}^l\right) \tag{49}$$

$$\Delta x_{ij}^{l} = W_{H}^{l} Concat_{k \epsilon 1,\ldots,H}\left(\Delta x_{ij}^{k,l}\right) \tag{50}$$

$$\Delta x_{i}^{l} = Aggregation_Sum_{j \epsilon N(i)}\left(\Delta x_{ij}^{l}\right) \tag{51}$$

$$x_{i}^{l+1} = x_{i}^{l} + \Delta x_{i}^{l} \tag{52}$$

where $x_i^l$, $x_j^l$, $\overrightarrow{\Delta x_{ij}^l}$, $\Delta x_{ij}^{k,l}$, $\Delta x_i^l$, and $x_i^{l+1} \epsilon R^{1\times3}$; $W_{x1}^l \epsilon R^{(d_k/2)\times d_k}$, $W_{x2}^l \epsilon R^{1\times(d_k/2)}$, $W_H^l \epsilon R^{1\times(H)}$, $b_{x1}^l \epsilon R^{(d_k/2)}$, and $b_{x2}^l \epsilon R^1$.

The Gate Block serves as a fundamental component for the residual connections in EGNN layers.

$$g = Sigmoid\left(Dropout\left(W_g Concat\left(\hat{h}_{new}, h_{old}, \hat{h}_{new} - h_{old}\right) + b_g\right)\right) \tag{53}$$

$$h_{new} = GraphNorm\left(g \odot \hat{h}_{new} + h_{old}\right) \tag{54}$$

where $h_{old}$, $\hat{h}_{new}$, $g$, and $h_{new} \epsilon R^{d_h}$; $W_g \epsilon R^{d_h \times d_h}$ and $b_g \epsilon R^{d_h}$ are learnable parameters from linear layers.

**Post-processing**

Since AutoLoop only predicts the backbone heavy atoms of the loop region, post-processing is utilized to add side chains and perform energy minimization to generate high-quality loop conformations. Therefore, we implemented a post-processing module that adds side chains and minimizes the energy of the loop conformation. For this purpose, we selected OpenMM[48], which is compatible with GPU acceleration, and used the ff14SB force field to optimize the predicted conformation. To balance computational cost and optimization quality, a high tolerance level was chosen to ensure efficient processing. In practical applications, the tolerance level can be adjusted automatically based on specific requirements. The implemented post-processing steps include:

1. Preprocessing: Utilizing PDBFixer to identify and fix missing residues, atoms, and hydrogen atoms in the input protein structure.
2. Force field application: Using the Amber14 force field to model the structure and add hydrogen atoms.
3. External force application: Introducing a custom external force to control the position deviation of key atoms in the loop region.

4. Molecular dynamics simulation: Performing simulations with the Langevin integrator on the CUDA platform to minimize energy and obtain optimized protein loop structures.

This post-processing approach ensures that the predicted loop conformations are refined to a high standard, suitable for further analysis and applications.

## Training protocol

In this study, we optimized the model using the Adam optimization algorithm. The parameters set included a batch size of 64, a learning rate of $1e^{-3}$, and a weight decay of $1e^{-5}$. The training is stopped if the loss on the validation set loss increases consecutively across 70 epochs. After that, the GT encoder and GVP encoder are believed to have partially captured the characterization of proteins adequately. Subsequently, we focused on training the autoregression module using previously prepared datasets. For efficiency and reliability throughout this training phase, we limited the number of generated atoms to 10 during training, while the full loop of atoms was generated during the inference stage. The loss function of the AutoLoop is based on the calculated backbone heavy atom RMSD between predicted loop conformations and the ground truth conformations. The training hyperparameters are maintained as previously described, with two adjustments: the learning rate is set to $1e^{-4}$ and the weight decay is canceled.

$$L_{rmsd} = RMSD\left(x_l^{pred}, x_l^{label}\right) = \sqrt{\frac{\sum_{n=1}^{N}\left(x_{l,n}^{pred} - x_{l,n}^{label}\right)^2}{N}} \tag{55}$$

where $N$ denotes the number of loop nodes and $n$ represents the index of the loop nodes.

## Evaluation methods

To assess AutoLoop's efficacy, we selected a range of established techniques for comparative analysis. We examined three main categories of protein loop modeling approaches:

1. **Knowledge-Based Approaches:** This category includes tools like FREAD, Prime, DaReUS-Loop, and LoopIng, which utilize historical data and recognized patterns to model protein loops.

2. **Ab Initio Approaches**: Techniques such as $D_ISG_{RO}$ , NGK, Rosetta-missing-loop (RML) and GalaxyLoop-PS2 fall into this category. These methods predict protein structures based on fundamental physical and chemical principles, without relying on existing template structures.
3. **Hybrid Approaches**: Represented by Sphinx, this approach merges elements of both knowledge-based and ab initio methodologies to model protein loops.
4. **Deep-learning Approaches**: Represented by AlphaFold2, AlphaFold3, RoseTTAFold and ColabFold, which predicts the whole protein conformation based on the sequence information.

It is noteworthy that NGK requires a preliminary loop conformation to function effectively. In scenarios where a loop conformation is absent, RML proves to be a practical alternative for its reconstruction. Following the default settings outlined in the tutorials for each method, we configured the output structure to generate one decoy. Despite the NGK tutorial recommending the generation of 500 decoys, this approach resulted in inferior performance compared to generating a single decoy, with the accuracy decrease from 1 decoy average RMSD 3.40 Å, median RMSD 1.88 Å to 500 decoys 3.51 Å, 2.52 Å and takes 21 days on 36-core CPU to complete prediction. Thus, the single-decoy approach (used in our study) provides better accuracy and practical feasibility, aligning with standard practices in loop modeling. Our study involved applying the methods of FREAD, Prime, DISGRO, NGK, RML, AlphaFold2, AlphaFold3, RoseTTAFold and ColabFold to the CASP15 dataset. Additionally, we examined the performance of LoopIng, DaReUS-Loop, Sphinx, GalaxyLoop-PS2, and NGK in predicting loop conformations within the HOMSTRAD dataset, drawing on data supplied by DaReUS-Loop.

**Computing resource**

AutoLoop was trained on 4 NVIDIA A100-SXM4-80GB and 64 cores Intel(R) Xeon(R) Platinum 8358P CPU @ 2.60GHz. For evaluation, AutoLoop was evaluated on a Tesla V100S GPU and a single-core Intel(R) Xeon(R) Gold 6240R CPU @ 2.40GHz CPU.

The conformation predicted by FREAD, Prime, $D_{I}SG_{RO}$ , NGK and RML on CASP15 was accomplished in parallel with 48 cores Intel(R) Xeon(R) Gold 6240R CPU @ 2.40GHz. AlphaFold2, RoseTTAFold and ColabFold was tested on a Tesla V100S and 20 cores Intel(R) Xeon(R) Gold 6240R CPU @ 2.40GHz CPU. AlphaFold3 was implemented through web-server. Loop conformations on the HOMSTRAD dataset predicted by LoopIng, DaReUS-Loop, Sphinx, GalaxyLoop-PS2, and NGK were provided by DaReUS-Loop.

## Data and Code Availability

The source code and testing datasets are available at https://zenodo.org/records/11293401.

## Supplementary Information

Figure S1 shows the comparison between the RMSDs of each sample predicted by AutoLoop and other tested methods. Figure S2 shows the comparison between the RMSDs of each sample predicted by AutoLoop_p and other tested methods. Figure S3 explains how the ligand and mutation residues impact the loop region, and analyze how the influence happens.

## Acknowledgments

This work was financially supported by National Natural Science Foundation of China (22377111).

## Author Contributions

T.W., C.H., T.H. and Y.K. designed the research study. T.W and X.Z. developed the method and wrote the code. L.W., O.Z., J.W., E.W., J.W., R.H., J.G., S.L., Q.S. and J. Y. performed the analysis. T.W., X.Z., C.H., T.H. and Y.K. wrote the paper. All authors read and approved the manuscript.